\documentclass[11pt]{article}
\usepackage{moriond,epsfig}

\def\be{\begin{equation}}
\def\ee{\end{equation}}
\def\bea{\begin{eqnarray}}
\def\eea{\end{eqnarray}}

\begin{document}
\vspace*{4cm}
\title{Constraints on high energy phenomena from low energy
nuclear physics\footnote{Presented at the XXIst Moriond Workshop
{\it Very High Energy
Phenomena in the Universe}, Jan. 20-27.2001, Les Acrs 1800, France.}}
 
\author{C. Hanhart}

\address{
Dept. of Physics, University of Washington,\\
Seattle, WA 98195-1560, USA \\
and\\
Institut f\"ur Kernphysik\\
Forschungszentrum J\"ulich\\
52428 J\"ulich, Germany} 

\maketitle\abstracts{
A procedure to derive bounds on coupling
strengths of exotic particles to nucleons
from the neutrino signal of supernovae is
outlined. The analysis is based on a model
independent calculation for the emissivities
for the exotic, detailed simulation for
the evolution of the early proto-neutron
star as well as a Likelihood analysis. 
As an example
we derive confidence levels for the upper bound
of the size of gravity only extra dimensions.}

In the aftermath of a core collapse supernova we find a system described
by extreme parameters: densities of several times nuclear matter
density with simultaneous appearance of temperatures of several tens
of MeV. Indeed, the matter is that dense and hot that even particles
interacting as weakly as neutrinos---the by far most efficient conventional cooling
mechanism---get trapped. Therefore, the cooling of the nascent proto--neutron star
(PNS) has to happen on diffusion times scales of tens of seconds. It is this
delayed cooling in the standard scenario that makes core collapse supernovae
one of the most efficient elementary particle physics labs possible: any additional
cooling mechanism will influence the time structure of the neutrino signal as
long as its interaction with the medium is not too strong that it gets trapped
as well~\cite{GGR}. 

\begin{figure}[t]
\begin{center}
\epsfig{file=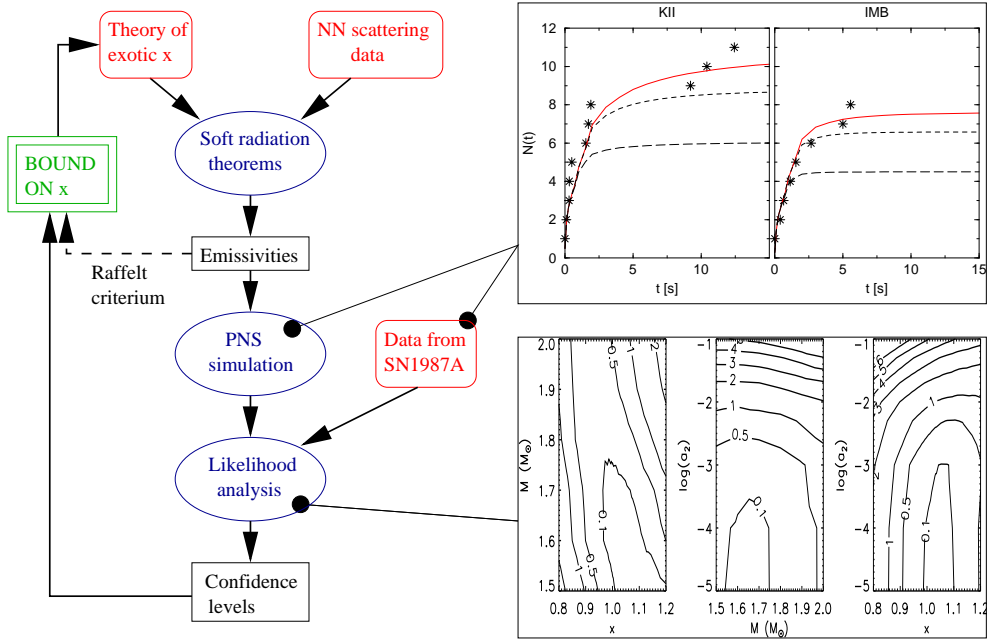, height=8.5cm}
\caption{\it{The structure of the analysis. Between
input (framed in rounded boxes), calculated quantities (boxes)
and analysis tools (ellipses) is explicitly distinguished. The
right column shows the comparison of the results of the
simulations to the measured integrated number
of counts (solid: $a_2$=0, short dashed: $a_2$=0.01, long dashed $a_2$=0.1)
 as well as different contours of the log--Likelihood function $q$. }}
\label{flussdia} 
\end{center}
\end{figure}

To make this kind of an argument more quantitative several ingredients are in
principle necessary: first one needs a reliable method to calculate the primary
emission rates of the exotic under discussion in the inner core. The calculated
emissivities are then to be fed into a simulation capable of describing the
first several tens of seconds of a nascent PNS. The results of these simulations
carried out for different coupling strengths of the additional cooling mechanism
are the to be compared with the data and analyzed with the appropriate statistical
method. This leads then to bounds on the coupling strength of the exotic studied.
The chain of arguments is illustrated in figure (\ref{flussdia}). Also indicated
in the diagram there is a possible short--cut, here labeled as Raffelt criterium.
This particular criterium~\cite{GGR} is one of a group of criteria meant to allow to read
a constraint off the emissivities directly. Although useful to get a very rough
estimate of the allowed strength of an exotic one should keep in mind that the 
emission rates of particles in the nuclear medium are strongly temperature and energy
dependent. The core temperature in the supernova, however, depends on whether or
not there is an additional cooling mechanism present. Thus it is difficult if not 
impossible to fix a fudicial temperature and density to be used in the criteria that
is valid for all kinds of particles (note that the temperature dependence of a 
particular production process depends on the energy/momentum dependence
of the fundamental coupling vertex of the exotic to the medium and thus depends
of the properties of the particle studied). Thus, to get a reliable bound, a 
detailed simulation is required after all.

Let us now go through the structure of the analysis in more detail: the first
step is the calculation of the emissivity of the exotic. In our work we follow
the method developed in ref. \cite{H1}. It is based on the observation that,
if the radiation is soft, the production rate can be expressed in terms of
the on--shell nucleon nucleon scattering data directly---the process
is dominated by radiation off external legs. Obviously this kinematic
requirement is best fulfilled in degenerate nuclear matter as it occurs in
the late cooling phase of a PNS. 
It was shown in ref. \cite{H1} that our model independent
approach leads to rates for neutrino--anti-neutrino bremsstrahlung that are
significantly smaller than those used so far. This finding was confirmed recently,
as reported in ref. \cite{dieperink}. The separation of scales is not that clear
when we move to the physical conditions present in the first 20 s after the collapse.
With temperatures of some tens of MeV and densities of several times nuclear matter
density, matter should be expected to be neither degenerate nor non--degenerate.
However, the investigations of ref. \cite{H2} suggest, that the relevant
expansion parameter, namely the radiation energy in units of the average 
nucleon--nucleon energy, still is smaller than one even here. 
Using this approach emissivities where already calculated 
for axions~\cite{H1} as well as gravition only extra dimensions (GODs)~\cite{H2}.
In addition the latter results where used as a basis for bounds derived on the
ground of cosmological arguments~\cite{cosmo,newraff}.

 Independent of
the production mechanism, it is always possible to cast the production rate in 
the following form
\begin{equation}
\frac{dE}{dt} = a_n \left(\frac{n_B}{n_0}\right)
\left(\frac{T}{10~{\rm MeV}}\right)^{p_n} \chi(X_n,X_p)
\quad {\rm MeV/baryon/s} \ .
\label{em-gr}
\end{equation}
To be concrete, let us concentrate on the production of gravitons in the 
presence of $n$ gravity only extra dimensions
(GODs)~\cite{gods}. There we find
$p_2=5.42$ , $a_2 = 5.1 \times 10^{4} R_2^2$, $p_3=6.5$, $a_3 = 1.4 \times
10^{16} R_3^3$. Here, $n_B$ is the baryon number density, $T$ is the
temperature. The size of the extra dimensions, $R_n$, is given in
mm. The emissivity depends weakly on the 
composition of matter through the function $\chi(X_n,X_p)$, defined in ref. \cite{H3}.

Once the emissivities are derived we have to quantify their influence on the
neutrino signal of SN1987A. The by far cleanest way to do so is to implement 
the emission rate of eq. (\ref{em-gr}) in a simulation of the early PNS.
For the present study we use the codes described in ref. \cite{P1}. The 
equilibrium diffusion approximation employed in the
numerical code provides a fair description of the total neutrino
luminosity and its time structure (c.f. figure (\ref{flussdia})) and thus
forms a sound basis for the present investigation. However, in order
to deduce the anti--neutrino spectrum an additional assumption is needed, namely
that the neutrino emission happens from a common neutrino sphere whose 
temperature is determined by the requirement for the optical depth to take
some fixed value. In order to minimize the influence of this assumption on 
the bound derived we studied a variety of values for the neutrino sphere temperature
around the central value. In addition, the mass of the PNS very strongly influences
the neutrino luminosity~\cite{P2}. Since we only know, that the baryonic
 mass is bounded in the
interval $1.4M_\odot \le M \le 2.0M_\odot ,$ we studied the whole mass range.

 Having neutrino emission rates at hand we now have to compare to the data
and develop a tool that quantifies the agreement. Thus we need to define a 
Likelihood function. Following ref. \cite{LL} we use
\begin{equation}
{\cal L}(\{ data\} |a_n,M,T_{\bar{\nu}_e},I) = 
\prod_D\left[ \prod_{i=1}^{N^{\rm tot}_D}\frac{dN_D(t_i^D)}{dt}\Delta t
\right] e^{-N_D} \ , 
\label{probdens}
\end{equation}
where $N_{D}$ is the total number of observed neutrino arrivals, and
$t_1^D,t_2^D,\ldots$ are the times at which neutrinos actually arrived in the
detector $D$.  $N_D(t)$ is the total number of neutrinos that the model with
assumptions $I$, PNS mass $M$, anti-neutrino temperature $T_{\bar{\nu}_e}$, and
exotic-coupling $a_n$ predicts will have arrived in the detector $D$ up until
the time $t$. $\Delta t$ can be any interval small enough that the probability
of detecting more than one count in any one bin can be taken to be
negligible. Ultimately it will be absorbed into an overall normalization
constant. 

The likelihood function provides us a with quantitative tool to
compare models of PNS evolution.  Since we treat not only the coupling
to extra dimensions, $a_n$, but also $M$ and $T_{{\bar \nu}_e}$, as
parameters, the likelihood function is a function in a
three-dimensional space. This function has a minimum at a baryonic
mass of $M=1.5 M_\odot$, $a_2=0$ and $T_{\bar{\nu}_e}=1.1 T_\nu^o$. 
$T^o_\nu$ denotes the
``reference anti-neutrino temperature'' defined above using the optical-depth
prescription. We can now assess all other models by looking at the log
of the ratio of their likelihood of a particular model to this ``most
likely'' model:
\begin{equation}
q(a_n,M,T_{\bar{\nu_e}})=-\log \left(\frac{{\cal L}(\{ data\}
|a_n,M,T_{\bar{\nu}_e},I)}{{\cal L}(\{ data\}
|0,1.5 M_\odot,1.1 T_\nu^o,I)}\right)
\label{eq:q}
\end{equation}
The function $q$ is then a function in this same three-dimensional
space. 

For the case of two GODs we discuss first the situation where the extra dimensions
have zero radius, and so $a_2=0$.  Contours of $q$ in the
resulting two-dimensional $x-M$ plane are shown in the left panel of
the contour plots in Fig.~(\ref{flussdia}). 
Here the horizontal axis is the temperature
normalized to the reference anti-neutrino temperature, $x=T_{{\bar
\nu}_e}/T_\nu^o$, which is varied to explore the sensitivity of the
results to deviations of the neutrino temperature from that obtained
using the optical depth prescription. 
 We see that although $M=1.5
M_\odot$, $T=1.1 T_\nu^o$ is indeed a minimum of the function $q$ it
is a rather shallow minimum: varying the neutron-star mass and the
anti-neutrino temperature over the range considered here does not
reduce the likelihood greatly. Ultimately this reflects the weakness
of the constraint that the SN 1987a data provides for these
parameters.

The situation is rather different as we move away from $a_2=0$. In the
middle panel we display similar likelihood contours in the $M-\log(a_2)$
plane for the case $T_{{\bar \nu}_e}=T^o_\nu$, while the right panel
shows contours in the $x-\log(a_2)$ plane for the baryonic mass 
$M=1.6 M_\odot$. These
panels show that the likelihood function decreases rapidly for $a_2
\ge 10^{-2}$---regardless of the values of the poorly-known parameters
$M$ and $T_{{\bar \nu}_e}$. Such large values of $a_2$ are essentially
two orders of magnitude less likely than the ``most likely''
model. This is a contrast to smaller values of $a_2$, where the
differences in likelihood are comparable to those seen in the
$M-T_{{\bar \nu}_e}$ plane. Thus statements about the most likely
value of $a_2$ in this smaller-$a_2$ regime will be sensitive to the
ill-constrained information on these neutron-star
parameters. Consequently we will not make any such statements
here. However, the two rightmost panels give us confidence in our
ability to derive a {\it bound} on $a_2$---as opposed to a most-likely
value---since it is clear that certain values of this coupling can be
well-excluded, completely independent of details of the PNS modeling.
To actually derive a bound 
we now integrate (or ``marginalize'') over all possible values
of $M$ and $T_{\bar{\nu}_e}$, using appropriate weight functions
(for details c.f. ref. \cite{H3}). 
We find that the possibility that there are two compact extra
dimensions with radii larger than 0.66 $\mu$m is excluded at the 95\%
confidence level---as is the possibility that there are three compact extra
dimensions larger than 0.8 nm~\cite{H3}.

To summarize, in this talk a scheme to deduce bounds with a well defined
statistical meaning for the coupling strength of exotic particles to nucleons
was presented. The method is quite general and it was demonstrated that the
bounds derived do not depend on parameters poorly constrained. However, improvements are
possible, and should be studied. For instance, many-body effects are expected
to modify the emissivity.
We are working on this problem.

{\bf Acknowledgements}: the results presented are based on a
fruitfull and enjoyable collaboration 
with D.~R.~Phillips, J.~A.~Pons, S.~Reddy and M.~J.~Savage---thanks.

\end{document}